\PassOptionsToPackage{table,dvipsnames}{xcolor}
\documentclass[a4paper,fleqn]{cas-dc}

\usepackage[numbers]{natbib}
\usepackage{xcolor}
\usepackage[most]{tcolorbox}
\usepackage{colortbl,hhline}
\usepackage{booktabs}
\usepackage{cellspace}
\usepackage{makecell}
\usepackage{subcaption}
\usepackage{graphicx}
\usepackage{wrapfig}
\usepackage{amsmath}
\usepackage{amsfonts}

\usepackage{pifont}

\usepackage[ruled,vlined]{algorithm2e}
\usepackage{algpseudocode}

\usepackage{float}
\usepackage{placeins}
\usepackage{afterpage}
\usepackage{flushend}

\usepackage{enumitem}
\usepackage{comment}

\usepackage{hyperref}
\usepackage[capitalize]{cleveref}
\crefformat{footnote}{#2\footnotemark[#1]#3}
\newcommand{\nba}{NBA}
\usepackage[separate-uncertainty = true, multi-part-units = repeat]{siunitx}

\usepackage{threeparttable}
\usepackage{adjustbox}
\usepackage{array}
\usepackage{longtable}
\usepackage{tabularx}
\newcolumntype{Y}{>{\raggedright\arraybackslash}X}
\usepackage{ltablex}
\usepackage{arydshln}
\usepackage{multirow}
\usepackage{lipsum}
\usepackage{tikz}
\usetikzlibrary{arrows.meta}
\keepXColumns

\newcommand{\srq}[1]{%
  \par\medskip
  \noindent\textbf{\textit{#1}}\enspace
}

\newcommand{\parg}[1]{%
  \par\medskip
  \noindent\textbf{\textit{#1}}\enspace
}

\newtcolorbox{GrayBox}[2][]{%
  enhanced,
  breakable,
  width=\columnwidth,
  colback=gray!10,
  colframe=black!70,
  arc=1mm,
  boxrule=0.5pt,
  left=1ex,
  right=1ex,
  top=1ex,
  bottom=1ex,
  before skip=2ex,
  after skip=3ex,
  title=#2,
  #1
}

\newcommand{\best}[1]{\cellcolor{gray!25}{#1}}
\newcommand{\nb}[1]{\cellcolor{white!20}{#1}}

\def\tsc#1{\csdef{#1}{\textsc{\lowercase{#1}}\xspace}}
\tsc{WGM}
\tsc{QE}

\begin{document}
\let\WriteBookmarks\relax

\shorttitle{GNNs for Code-to-Architecture Mapping}

\shortauthors{Jabeen et al.}

\title[mode = title]{When Does the Dependency Graph Help Code-to-Architecture Mapping?}

\author[1,3]{Rakhshanda Jabeen}[orcid=0009-0002-0430-8179]
\cormark[1]
\ead{rakhshanda.jabeen@electroluxprofessional.com}
\credit{Conceptualization, Methodology, Software, Investigation, Writing (original draft, review \& editing)}

\affiliation[1]{organization={Department of Computer Science},
            addressline={Linnaeus University},
            city={Växjö},
            country={Sweden}}

\affiliation[2]{organization={Department of Mathematics},
            addressline={Linnaeus University},
            city={Växjö},
            country={Sweden}}

\affiliation[3]{organization={Electrolux Professional AB},
            city={Ljungby},
            country={Sweden}}

\author[1]{Morgan Ericsson}[orcid=0000-0003-1173-5187]
\ead{morgan.ericsson@lnu.se}
\credit{Conceptualization, Software review, Supervision, Writing (review \& editing)}

\author[2]{Jonas Nordqvist}[orcid=0000-0002-0510-6782]
\ead{jonas.nordqvist@lnu.se}
\credit{Supervision, Writing (review \& editing)}

\author[1]{Anna Wingkvist}[orcid=0000-0002-0835-823X]
\ead{anna.wingkvist@lnu.se}
\credit{Supervision, Writing (review \& editing)}

\cortext[1]{Corresponding author}

\begin{abstract}
Architectural technical debt arises when a system's implementation drifts from its intended modular decomposition, while the mapping between source files and architectural modules is missing, incomplete, or costly to maintain manually. We formulate code-to-architecture mapping as semi-supervised node classification over a directed multi-relational dependency graph, with files as nodes, typed dependencies as edges, and lexical features derived from code identifiers and directory layout. Starting from a small labeled sample, an iterative self-training procedure expands the mapping to the remaining files. We compare a graph convolutional network (GCN), a graph attention network (GAT), and a graph-free multi-layer perceptron (MLP), using identical features and training procedures for the learned models, against a Naive Bayes reference. Across 100 runs on ten open-source Java and C/C++ systems, all three neural models outperform the reference on both F1 metrics. GAT achieves the highest averages of system-level median macro and micro F1, at 83.5 and 92.2, respectively, whereas MLP remains competitive or superior on several systems. On Chrome, the largest system with 18,343 files, GAT and GCN exceed MLP by 22 and 19 macro-F1 points, respectively, and have substantially fewer persistently misclassified files. Connectivity and homophily are associated with where neural message passing helps, while multi-relational aggregation outperforms a single merged edge type on every system for both GCN and GAT. A follow-up study finds that several model rankings depend on the training seed, qualifying smaller advantages without removing the large graph benefit observed on Chrome.
\end{abstract}


\maketitle

\section{Introduction\label{sec:intro}}

Software architecture defines a system's high-level structure and the dependencies among its principal elements, and it provides the basis for reasoning about maintainability, security, and other quality attributes~\cite{clements2003}. That reasoning requires the implementation to stay aligned with the intended design. When it does not, the gap is architectural technical debt~\cite{kruchten2012technicaldebt}: cheap to introduce, but diffuse and relational rather than local to a single file, and therefore expensive to detect and repay once it accumulates. Large language models (LLMs) are increasingly used to generate or assist with code commits~\cite{ehsani2025fastercode}, motivating renewed attention to whether local code changes remain consistent with a system's module structure. An empirical study of commercial AI code generators found clones between generated functions and existing open-source code~\cite{wu2025clones}. These observations motivate conformance checking, but do not establish that LLM-generated code causes architectural erosion. Architectural erosion more generally concerns drift between the intended design and its implementation~\cite{de2012controlling}.

Detecting this drift requires comparing the intended architecture against the implemented one. Static Architecture Conformance Checking (SACC) addresses this comparison, and reflexion modeling is a SACC technique that classifies relations between the intended and implemented architectures as convergent, divergent, or absent~\cite{reflexion1,reflexion2}. However, reflexion modeling depends on a mapping between source code entities and architectural modules, and such mappings are often missing, incomplete, or expensive to maintain manually~\cite{ali2018architecture}. Even for medium-sized systems, recovering or maintaining the architectural decomposition can require substantial expert effort~\cite{garcia2013}. Automated code-to-architecture (C2A) mapping is therefore an important prerequisite for practical conformance checking.

Prior work on C2A mapping has explored supervised and recommendation-based approaches. Supervised approaches compute \emph{attraction values} that quantify the association strength between source code entities and architectural modules~\cite{christl2007automated, bittencourt_IR}. Olsson et al.~\cite{olsson2022automatically} proposed a semi-supervised Naive Bayes mapper that combines syntactic and semantic information to assign source code entities to architectural modules. InMap~\cite{InMap} took a recommendation-based view instead, using information retrieval over the codebase to suggest candidate mappings with minimal architectural documentation and no training data.

A separate line of work targets architecture recovery, where the decomposition itself must be discovered rather than assumed. Recent recovery approaches include graph neural networks (GNNs) combined with contrastive learning and autoencoders~\cite{gaer,arl_2026}, along with language models applied directly to source code for recovering design structures~\cite{amalfitano2025automated}, for agent-based recovery of large legacy systems~\cite{archagent}, and for extracting architectural entities to support traceability~\cite{exarch}. The distinction from C2A mapping matters here: the module set is unknown in recovery, whereas in mapping it is given by the intended architecture, and the task is to assign each file to one of those modules. It also matters for evaluation, since many recovery benchmarks are open source, and a decomposition reproduced by a pretrained model may reflect code understanding, prior exposure to the system, or both. Mapping methods that learn from a system's own structure and can be evaluated against a known ground truth therefore remain important.

In this research, we study C2A mapping\footnote{\label{fn:code}Source code is available at \url{https://github.com/sse-lnu/GCN-SAEroCon25}.} as semi-supervised node classification over a multi-relational directed dependency graph. Source files are nodes, typed static dependencies between them are edges, and node features encode code identifiers via Word2Vec embeddings together with directory context. A small set of labeled files is expanded into a complete mapping through iterative self-training. We compare three encoders under the same features and training procedure: a graph convolutional network (GCN), a graph attention network (GAT), and a graph-free multi-layer perceptron (MLP). Holding the feature representation, the classifier head, and the self-training loop fixed is intended to isolate what the dependency graph contributes beyond the node features alone, though the encoders themselves differ in architecture beyond the presence or absence of message passing.

We evaluate the approach on ten open-source systems written in Java and C/C++, reporting mapping accuracy and its variation across runs. This paper extends our earlier study~\cite{jabeen2025graph} in five respects: it broadens the benchmark by adding C/C++ systems; replaces the earlier bag-of-words representation of code identifiers with Word2Vec embeddings; adds GAT to assess the effect of attention-based aggregation on the mapping task; introduces an MLP baseline to assess the contribution of dependency information; and corrects a label-assignment defect in the earlier self-training pipeline. We further evaluate the corrected self-training procedure, conduct new ablation studies of model depth and edge-aggregation direction, and examine the sensitivity of selected comparisons to the training seed.

\noindent Our study is guided by the following research questions:

\begin{itemize}[leftmargin=1em]
    \item \textbf{RQ1.} How effectively does GNN-based C2A mapping perform relative to non-graph baselines, and how stable is its performance across runs?
     \item \textbf{RQ2.} What does the dependency graph contribute beyond textual features under an identical training procedure, and which system properties are associated with where it helps?
    \item \textbf{RQ3.} How sensitive is the approach to modeling dependency types separately, aggregation depth, and edge direction?
\end{itemize}

\section{Related Work\label{sec:work}}

\citet{tzerpos1997orphan} introduced the orphan adoption problem, in which source code entities that are not yet part of the documented architecture are incrementally assigned to modules. \citet{christl2007automated} built on this idea with the Human-Guided Clustering Method (HuGMe), a semi-automatic approach that recommends mappings through attraction functions and leaves the final decision to the architect. \citet{bittencourt_IR} studied attraction functions derived from information retrieval, where an entity is assigned to the module whose textual content it most resembles. These approaches rely on two signals that persist in later work: lexical similarity between an entity and a module and dependency proximity to entities that are already mapped, but they combine them through fixed, manually specified rules.

Subsequent work replaced these rules with learned models. Olsson et al.~\cite{olsson2022automatically} proposed a semi-supervised Naive Bayes mapper that combines lexical information from source code with dependency-derived information through Concrete Dependency Abstraction, and reported that 15--20\% of the entities must be mapped beforehand to reach acceptable performance. \citet{florean2021textclassifierscomparison} compared text classifiers for the same task and showed that both the classifier and the text representation affect mapping quality. InMap~\cite{InMap} instead ranks candidate mappings interactively, using natural language module descriptions together with source code, which removes the need for a pre-mapped training set. In all of these, dependencies enter the model only after being flattened into features attached to individual entities, so the relational structure of the system is consumed rather than represented.

Language models have recently been applied to the same task. \citet{rukmono2024deductive} use chain-of-thought prompting with GPT-4 to deductively assign implementation units to a predefined reference architecture. \citet{johansson2024mapping} evaluate language model embeddings alongside TF-IDF and GloVe as input to method-level classifiers. Follow-up work links variation across projects most strongly to class separability and density in the embedding space~\cite{johansson2026mapping}. A parallel line applies language models to architecture recovery, where the module set is inferred rather than given~\cite{amalfitano2025automated,archagent}. Pretrained models capture semantic cues that engineered text features miss. ArchAgent, for example, also incorporates reference graphs and related-file context when summarizing code. Our approach instead models dependency types directly through relation-specific message passing.

Graph representation learning uses dependencies as structure rather than as features. NEGAR~\cite{negar} showed that random-walk embeddings over the dependency graph, followed by clustering, recover meaningful architectural structure. GAER~\cite{gaer} extended this to heterogeneous graphs, comparing GCN and GAT encoders trained within a graph autoencoder and modeling each dependency type as a separate relation, and \cite{arl_2026} reports related results using contrastive objectives. Both are unsupervised recovery methods, so the module set is discovered by clustering rather than fixed in advance as it is in mapping. Our earlier work~\cite{jabeen2025graph} took this step and evaluated GCN-based models on homogeneous and dependency-type-aware graphs, showing that a small labeled set can be expanded through self-training. That study was limited to Java systems of moderate size, used a single graph-based encoder family, and did not isolate how much of the observed performance came from the dependency graph as opposed to the node features.

The present work addresses these limitations. We evaluate GNN encoders together with a graph-free neural baseline that receives identical features and an identical self-training loop, to examine the contribution of dependency information under a common input representation and self-training schedule. Differences in encoder depth, dropout, and parameter count remain potential confounders.

\section{Methodology\label{sec:method}}
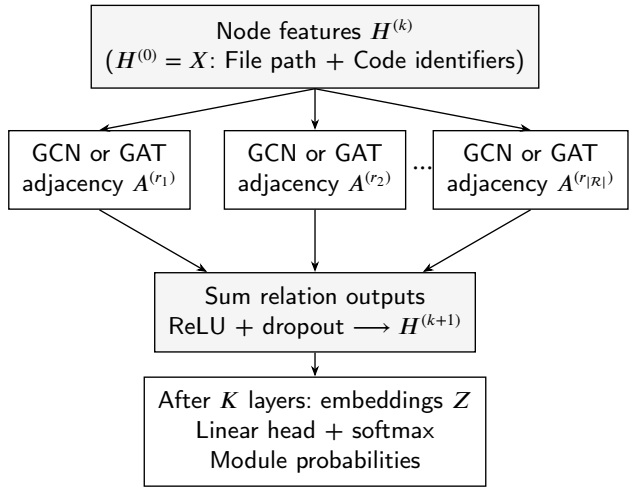
\begin{figure}[pos=!htbp]
\centering
\resizebox{0.98\columnwidth}{!}{%
\begin{tikzpicture}[font=\small,
  block/.style={draw,align=center,minimum height=0.8cm,inner sep=5pt},
  >={Stealth[length=4pt,width=3pt]},
  every path/.style={line width=0.5pt}]
\node[block,fill=gray!8] (input) at (0,0)
  {Node features $H^{(k)}$\\($H^{(0)}=X$: File path + Code identifiers)};
\node[block] (r1) at (-2.8,-1.6)
  {GCN or GAT\\adjacency $A^{(r_1)}$};
\node[block] (r2) at (0,-1.6)
  {GCN or GAT\\adjacency $A^{(r_2)}$};
\node[block] (rn) at (2.8,-1.6)
  {GCN or GAT\\adjacency $A^{(r_{|\mathcal{R}|})}$};
\draw[->] (input.south) -- (r1.north);
\draw[->] (input.south) -- (r2.north);
\draw[->] (input.south) -- (rn.north);
\node at (1.4,-1.6) {$\cdots$};
\node[block,fill=gray!8] (sum) at (0,-3.45)
  {Sum relation outputs\\ReLU + dropout $\longrightarrow H^{(k+1)}$};
\draw[->] (r1.south) -- ([xshift=-1.4cm]sum.north);
\draw[->] (r2.south) -- (sum.north);
\draw[->] (rn.south) -- ([xshift=1.4cm]sum.north);
\node[block] (head) at (0,-5.0)
  {After $K$ layers: embeddings $Z$\\Linear head + softmax\\Module probabilities};
\draw[->] (sum.south) -- (head.north);
\end{tikzpicture}%
}
\caption{Overview of the heterogeneous GNN encoder. Node features $X$ (file-path indicator + Word2Vec identifiers) and relation-specific adjacency matrices $A_\mathcal{R}$ are passed through a relation-specific encoder (GCN or GAT), whose outputs are summed and passed through a non-linear activation function (ReLU) to produce embeddings $Z\in \mathbb{R}^{|\mathcal{F}| \times d_h}$, where $d_h$ is the number of hidden channels. A linear layer maps $Z$ to per-module logits, and a final softmax produces the module probabilities used for classification and pseudo-labeling.\label{fig:pipeline}}
\end{figure}

This section presents our approach to C2A mapping as semi-supervised node classification over a directed multi-relational dependency graph. We first formulate the mapping task, then describe the graph and feature construction, introduce the encoders compared in this study, and finally the self-training procedure used to grow an initial labeling into a full mapping. Figure~\ref{fig:pipeline} illustrates the heterogeneous GNN encoder at the core of this pipeline.

\subsection{Problem Formulation \label{sec:problem}}
 
Given a software system with a set of files $\mathcal{F}$ and a fixed set of architectural modules $\mathcal{C}$, a C2A mapping assigns each file $i \in \mathcal{F}$ to a module $y_i \in \mathcal{C}$. A small subset $L \subset \mathcal{F}$ is labeled with its ground-truth module; the remaining files $U = \mathcal{F} \setminus L$ are unlabeled, and the task is to recover the assignment for every file in $U$. We consider a transductive setting in which the lexical node features and dependency structure of all files are available during training, while module labels are available only for files in $L$.

Two sources of information are available for each file. The first is intrinsic and consists of what the file itself declares (identifiers) and where it sits in the source tree (directory structure), and it is available for every file regardless of how the file is connected to the rest of the system. The second is relational: the representations of its neighbors in the dependency graph, propagated through message passing and available only where the file has neighbors. We frame the task as representation learning so the two sources can be treated separately: every encoder maps a file $i$ to an embedding $z_i$, and a shared classification head maps $z_i$ to a module, so the encoders differ in what evidence enters the encoder to generate $z_i$ and not in how it is used.

\subsection{Graph Construction}

We represent a software system as a heterogeneous graph $G = (X, A_\mathcal{R})$ over its files, following the construction introduced for unsupervised recovery in GAER~\cite{gaer}. Code identifiers and directory paths are file-level properties and are encoded in the feature matrix $X$. Static dependencies are file-to-file relations, each belonging to a dependency type $r$ drawn from the set of relation types $\mathcal{R}$, and encoded as edges in the multi-relational adjacency $A_\mathcal{R} = \{A^{(r)}\}_{r \in \mathcal{R}}$.

\subsubsection{Node Features \label{sec:features}}
The intrinsic evidence for a file comes from two conventions developers follow when they write and organize code. Identifiers name the concepts a file works with, so class, method, and variable names carry the vocabulary of the domain the file serves and, with it, an indication of the responsibility it implements~\cite{sarif}. Directory paths record how developers chose to group files, and that grouping frequently tracks the subsystem decomposition the architecture describes~\cite{garcia2013}. The two are complementary where structure alone is ambiguous, since a widely used utility file may depend on code throughout the system while its name and its location still indicate what it is for. Both signals are available after identifier extraction, without requiring additional dependency resolution during feature construction.

We normalize code identifiers by splitting on \emph{camelCase}, \emph{PascalCase}, and \emph{snake\_case} boundaries and removing low-information tokens (stopwords, very short tokens, tokens appearing in over 90\% of files), then train a 128-dimensional skip-gram Word2Vec~\cite{word2vec} embedding over the remaining tokens for 20 epochs with a context window of 5, a minimum token count of 5, and a vocabulary-pruning limit of 2,000 during vocabulary construction. We use one document per file, a fixed stopword list, and a single Word2Vec worker with a fixed seed. Let $T_i$ denote the resulting sequence of normalized tokens for file $i$; its textual representation is the $\ell_2$-normalized average of the Word2Vec embeddings of the tokens in $T_i$. File paths are tokenized separately and represented as a binary bag-of-words vector. Common leading directories and the generic components \texttt{src}, \texttt{main}, and \texttt{java} are removed. Java files use their remaining directory tokens, falling back to the file name when none remain; C/C++ files also include base-name tokens. Tokens are lowercased and vectorized using the default word-token pattern of \texttt{CountVectorizer}. For file $i$, the two are concatenated:
\begin{equation}
x_i = \big[\text{Binary(FilePath)}_i \,\big \Vert\, \text{Word2Vec}(T_i)\big].
\label{eq:features}
\end{equation}

Stacking $x_i$ over all files gives the feature matrix $X$. Every encoder in this study receives the same $X$, so the intrinsic evidence available to each of them is identical by construction, although the encoders also differ in depth and parameter count.
 
\subsubsection{Static Dependencies (Edges) \label{sec:graph}}
Files that implement related functionality tend to depend on one another directly or through short dependency paths, which is what makes the dependency graph informative for architecture recovery~\cite{structsar}. Dependency types are not interchangeable in this respect, and call edges have been argued to indicate tighter coupling than type-usage edges~\cite{negar,edge_weights}. Collapsing them into one adjacency matrix removes the distinction between dependency types, so we keep one adjacency matrix per type and let the encoder learn a separate transformation for each. Section~\ref{sec:rq3} evaluates this relation-typed design against a homogeneous graph for both GNN encoders.

We extract dependencies with the multilingual static-analysis tool \emph{Depends}~\cite{depends}, which resolves the typed relations listed in Table~\ref{tab:relation_types} between code entities across our Java and C/C++ systems. Because the mapping task operates at file granularity, member-level dependencies such as a call between two methods are aggregated into edges between their enclosing files, and repeated dependencies of the same type between the same ordered pair collapse into a single edge, so $A^{(r)}$ records whether a relation holds and not how often it occurs.

\begin{table}[pos=htbp]
\centering
\caption {Static dependencies used in the software graph. \label{tab:relation_types}}
\renewcommand{\arraystretch}{1.1}

\begin{adjustbox}{max width=\columnwidth}
\begin{tabular}{ll}
\toprule
\textbf{Dependency} & \textbf{Description} \\
\midrule
\texttt{Annotation} & Annotation relation to an annotated entity. \\
\texttt{Call} & Function or method invocation. \\
\texttt{Cast} & Conversion of an entity to another type. \\
\texttt{Contain} & Definition of a variable or field of another type. \\
\texttt{Create} & Creation of an instance or program entity. \\
\texttt{Extend} & Inheritance or parent--child type relation. \\
\texttt{Implement} & A class implements an interface or abstraction. \\
\texttt{Import} & A file imports other classes, methods, or variables. \\
\texttt{Parameter} & Use of a type as a method or function parameter. \\
\texttt{Return} & Use of a type as a return type. \\
\texttt{Throw} & A type is used as an exception. \\
\texttt{Use} & Use or assignment of variables. \\
\bottomrule
\end{tabular}
\end{adjustbox}
\raggedright\footnotesize \texttt{Annotation} and \texttt{Throw} are not observed in the C/C++ systems.
\end{table}

A static dependency is asymmetric, and $A^{(r)}$ preserves that direction: an edge points from the file where it originates to the file it refers to. For file $i$ under relation $r$, we distinguish its in-neighborhood, the files that depend on it, from its out-neighborhood, the files it depends on. A file with a large in-neighborhood and a small out-neighborhood provides functionality, the profile of an interface or utility, and its clients may be spread across many modules. A file with the opposite profile consumes functionality, like orchestration or entry-point code, and what it draws on says more about what it coordinates than about which module it belongs to. Which of the two an encoder aggregates over, or whether it discards the direction altogether and treats the edge as undirected, is therefore a modeling choice rather than a fixed property of the graph; Section~\ref{sec:encoders} defines it as a parameter of message passing, and Section~\ref{sec:rq3} tests its impact on mapping.

\subsection{Representation Learning and Mapping \label{sec:encoders}}
We compare three encoders that operate on the same node features $X$. The graph-based models, GCN and GAT, use the relation-specific adjacency $A_\mathcal{R}$ to aggregate information from neighboring files, while the non-graph baseline uses $X$ alone. All models take $h_i^{(0)} = x_i$ as input.
 
The GNNs stack $K$ message-passing layers to produce $z_i = h_i^{(K)}$, while the MLP uses one hidden layer. Each encoder feeds its embedding into a linear classification head followed by softmax to obtain module probabilities. The encoders differ in their use of neighbors, aggregation weights, depth, and parameter count.

\subsubsection{Graph Neural Networks (GNNs)}
The GNNs update the representation of each file by aggregating messages from its neighbors under each dependency type $r$, a mechanism known as message passing. The neighborhood $\mathcal{N}_i^{r,\pi}$ is determined by relation $r$ and propagation scheme $\pi$, as introduced in Section~\ref{sec:graph}. Forward propagation is source-to-target: a file aggregates from files that depend on it. Reverse propagation aggregates from files it depends on, while undirected propagation adds each edge in both directions.
 
A relation- and layer-specific transformation $W^{r,(k)}$ is applied to each neighbor's message, so different dependency types, such as calls and inheritance, contribute through separate learned transformations. A $K$-layer encoder incorporates information from up to $K$ dependency steps away. We add self-loops so each file's own features contribute to its update, even when it has no other neighbors. The neighborhoods in Equations~\ref{eq:gcn} and~\ref{eq:gat} include these self-loops.

\paragraph{GCN:} We use a relation-specific variant of the graph convolutional layer~\cite{gcn}, which updates each file's representation by aggregating its neighbors' messages, weighted by the inverse square root of the product of the endpoint degrees under each relation:
\begin{equation}
\label{eq:gcn}
h_i^{(k+1)} = \text{ReLU}\Big(\sum_{r \in \mathcal{R}} \sum_{j \in \mathcal{N}_i^{r,\pi}} \frac{1}{\sqrt{d_i^r d_j^r}} \, W^{r,(k)} h_j^{(k)}\Big),
\end{equation}
where $W^{r,(k)}$ is the relation- and layer-specific transformation, and $d_i^r$ and $d_j^r$ denote the sizes of $\mathcal{N}_i^{r,\pi}$ and $\mathcal{N}_j^{r,\pi}$, respectively. These degrees follow the propagation scheme $\pi$: incoming degrees for forward propagation and outgoing degrees for reverse propagation, both including self-loops. The implementation also learns a bias per relation and layer, omitted here for readability. ReLU~\cite{relu} is the non-linear activation. Degree normalization reduces the influence of high-degree files, such as widely used base types, on the aggregation.

\paragraph{GAT:} We use a relation-specific variant of the graph attention layer~\cite{gat}, which replaces fixed degree-based weighting with learned attention coefficients. For each relation $r$, the features of file $i$ and its neighbors are projected using $W^{r,(k)}$ and used to compute attention scores. A softmax over $\mathcal{N}_i^{r,\pi}$ normalizes these scores to obtain the attention weights $\alpha_{ij}^r$. The layer update for file $i$ is:
\begin{equation}
\label{eq:gat}
h_i^{(k+1)} = \text{ReLU}\Big(\sum_{r \in \mathcal{R}} \sum_{j \in \mathcal{N}_i^{r,\pi}} \alpha_{ij}^r \, W^{r,(k)} h_j^{(k)}\Big).
\end{equation}
Equation~\ref{eq:gat} shows a single head; the implementation concatenates four heads per relation before summing relation outputs and applying ReLU. Learned attention weights increase flexibility but introduce additional parameters. Changing the propagation scheme $\pi$ changes both the aggregated neighbors and their degree-normalization weights in GCN, whereas in GAT it changes the neighborhood over which attention is normalized.
 
\subsubsection{Multi-layer Perceptron (MLP)}
To assess the contribution of graph-neighborhood information, we use an MLP that operates on $x_i$ alone and does not aggregate messages from other files. For each file $i$, the layer update rule is defined as:
\begin{equation}
\label{eq:mlp}
h_i^{(k+1)} = \text{ReLU}\big(W^{(k)} h_i^{(k)} + b^{(k)}\big).
\end{equation}
Unlike the GNNs, whose representations incorporate information from neighboring files over multiple dependency hops, the MLP constructs each representation from the file's own node features only. Performance differences between the MLP and the GNNs are therefore consistent with a contribution of dependency-graph information, but do not isolate graph aggregation, since the encoders also differ in depth, dropout, and parameter count, with GAT introducing additional attention parameters.

\subsubsection{Classification head} All three encoders end in the same kind of output stage. The file embedding $z_i \in \mathbb{R}^{d_h}$, which is the output of the last message-passing layer for GCN and GAT and of the hidden layer for the MLP, is passed to a single linear layer that maps it to one score per architectural module. A softmax then converts these scores into a probability distribution over the modules:
\begin{equation}
\label{eq:head}
\begin{aligned}
p_i &= \text{softmax}\big(W_c z_i + b_c\big), \\
\hat{y}_i &= \arg\max_m p_{i,m}, \qquad \text{confidence}_i = \max_m p_{i,m},
\end{aligned}
\end{equation}
where $W_c \in \mathbb{R}^{|\mathcal{C}| \times d_h}$, $b_c \in \mathbb{R}^{|\mathcal{C}|}$, and $|\mathcal{C}|$ is the number of architectural modules. The module with the highest probability, $\hat{y}_i$, is the model's prediction for file $i$, and the self-training procedure in Section~\ref{sec:iterative} compares its probability, $\text{confidence}_i$, against the threshold $\tau$. The same softmax output enters the cross-entropy loss during training.

\subsection{Iterative Learning Framework \label{sec:iterative}}
GCN and GAT receive the same dependency graph and node features, differing only in how they aggregate and weight messages from neighboring files; the MLP uses the node features alone. Despite these differences, all three models follow the same self-training procedure, in which the confidence from Equation~\ref{eq:head} decides which files are pseudo-labeled.

Training begins with a fixed set of labeled files $L$, while the remaining files form the unlabeled set $U$. After an initial warm-up phase on $L$, each round scores every file. Any pseudo-label that the current prediction contradicts with confidence above the threshold $\tau$ is revised. Files from $U$ whose confidence exceeds $\tau$ are promoted into the mapped set $\mathcal{M}$, which initially equals $L$, with their predicted module as label. Ground-truth seed labels never change. The classifier is then trained further on $\mathcal{M}$ using full-batch cross-entropy, retaining the model parameters and Adam optimizer state across rounds. The procedure stops without further training when no new file exceeds $\tau$, or after $R$ rounds. Algorithm~\ref{alg:iter} summarizes this loop, identical across encoders except for the input representation and forward computation.
\begin{algorithm}
\small
\caption{Iterative Mapping Algorithm.\label{alg:iter}}
\KwIn{Input representation $\mathcal{I}$, files $\mathcal{F}$, labeled files $L$, confidence threshold $\tau$, warm-up epochs $E_w$, self-training epochs $E_s$, max rounds $R$}
\KwOut{Trained classifier $f$, mapped files $\mathcal{M}$}
$\mathcal{M} \gets L$;\ $U \gets \mathcal{F} \setminus L$\;
Train $f$ on $(\mathcal{I}, \mathcal{M})$ for $E_w$ epochs\;
\For{$k \gets 1$ \KwTo $R$}{
    Score all files in $\mathcal{F}$ with $f$ using $\mathcal{I}$\;
    Revise any pseudo-label in $\mathcal{M} \setminus L$ for which $f$ predicts a different module with confidence $> \tau$\;
    $\mathcal{P} \gets \{o \in U \mid \text{confidence}(o) > \tau\}$\;
    \If{$\mathcal{P} = \emptyset$}{\textbf{break}\;}
    Assign each $o \in \mathcal{P}$ its current highest-probability module as a pseudo-label\;
    $\mathcal{M} \gets \mathcal{M} \cup \mathcal{P}$;\ $U \gets U \setminus \mathcal{P}$\;
    Retrain $f$ on $(\mathcal{I}, \mathcal{M})$ for $E_s$ epochs\;
}
\Return{$f$, $\mathcal{M}$}\;
\end{algorithm}
\section{Experiments \label{sec:experiments}}
This section describes the experimental design used to answer the three research questions: the benchmark systems, configuration, baseline, and evaluation metrics.
\subsection{Benchmark Systems}
We use existing ground-truth (GT) mappings rather than constructing new ones, since a reliable C2A mapping requires substantial architectural knowledge, manual effort, and often expert validation~\cite{garcia2013}. Table~\ref{table:dataset_details} lists the ten Java and C/C++ systems in our evaluation.

Alongside system sizes, Table~\ref{table:dataset_details} reports graph properties, including average degree and file-level dependency homophily $\mathcal{H}$, the proportion of edges whose two files belong to the same GT module:
\begin{equation}
\mathcal{H} = \frac{|\{(i,j) \in \bar{E} : y_i = y_j\}|}{|\bar{E}|},
\label{eq:homophily}
\end{equation}
where $\bar{E}$ is the set of merged edges, containing $(i,j)$ if file $i$ depends on file $j$ under at least one dependency type. Homophily ranges from 0 (all dependencies cross modules) to 1 (all dependencies remain within modules).

\begin{table}[pos=h]
\centering
\caption{Properties of the software systems$^{\dagger}$. $|\mathcal{C}|$ denotes the number of modules in the GT architecture, edges $E$ are distinct typed dependencies between files, and the average typed out-degree is computed as $\langle k \rangle = |E|/|\text{Files}|$. \label{table:dataset_details}}
\renewcommand{\arraystretch}{1.1}
 
\begin{adjustbox}{max width=\columnwidth}
\begin{tabular}{c l l r r r c c}
\toprule
& System & Version
& $|\mathcal{C}|$
& $|\text{Files}|$
& $|E|$
& $\mathcal{H}$
& $\boldsymbol{\langle k \rangle}$\\
\midrule
 
\multirow{6}{*}{\rotatebox[origin=c]{90}{Java}}
&\href{http://argouml.tigris.org/}{A.UML}
& r13713  & 14 & 766 & 9\,491 & 0.55 & 12.4\\
 
&\href{https://commons.apache.org/proper/commons-imaging/}{C.Img}
& v1.0a2  & 21 & 329 & 4\,561 & 0.61 & 13.9\\
 
&\href{https://jabref.org/}{JabRef}
& v3.7  & 6 & 1\,180 & 17\,557 & 0.58 & 14.9\\
 
&\href{https://lucene.apache.org/}{Lucene}
& r1075001  & 7 & 1\,054 & 18\,595 & 0.67 & 17.6 \\
 
& \href{https://www.sweethome3d.com/}{SH-3D}
& r002382  & 9  & 167    & 3\,912   & 0.36 & 23.4 \\
 
&\href{https://TeamMatesv4.appspot.com/web/front/home}{T.Mates}
& v5.11  & 15 & 778 & 17\,818 & 0.34 & 22.9\\
 
\cmidrule{2-8}
\multirow{2}{*}{\rotatebox[origin=c]{90}{C}}
&\href{https://www.gnu.org/software/bash/}{Bash}
& 4.2  & 13 & 292 & 1\,222 & 0.60 & 4.2\\

&\href{https://gitee.com/openharmony/drivers_framework}{HDF}
& 0e196f  & 9 & 153 & 1\,239 & 0.35 & 8.1\\
\cmidrule{2-8}
\multirow{2}{*}{\rotatebox[origin=c]{90}{C++}}
&\href{https://www.chromium.org/}{Chrome}
& svn-171054  & 69 & 18\,343 & 284\,435 & 0.58 & 15.5\\

&\href{https://gitee.com/openharmony/distributed_camera}{HDC}
& 46ff87  & 11 & 207 & 1\,117 & 0.63 & 5.4\\
\bottomrule
 
\end{tabular}
\end{adjustbox}
\raggedright\footnotesize $\dagger$The file, module, and edge counts reflect the data used in our experiments, which for some systems omits a small number of files or edges from the raw pre-processed dataset.
\end{table}

The Java systems are drawn from the same source as our earlier work~\cite{jabeen2025graph}, assembled by Olsson et al.~\cite{olsson2022automatically} from systems previously studied at SAEroCon and by Brunet et al.~\cite{Brunet_gt} and Lenhard et al.~\cite{lenhard2019exploring}, with mappings distributed through the s4rdm3x GitHub repository\footnote{\href{https://github.com/h0bb3/s4rdm3x}{https://github.com/h0bb3/s4rdm3x}}.
 
For Bash, we use the C2A mapping curated by Garcia et al.~\cite{garcia2013}, produced by deriving an initial decomposition from documentation and source code and then validating it with system experts. The same line of work provides the GT mapping for Chrome, established through roughly two years of collaboration with the system's developers~\cite{deps_impact}.
 
The GT mappings for HDF and HDC were developed in collaboration with industry partners and are included in the SARIF benchmark studied by Zhang et al.~\cite{sarif}, distributed through the SARIF replication package\footnote{\href{https://github.com/anonymous2f4a9d/SARIF_FSE23/tree/main}{https://github.com/anonymous2f4a9d/SARIF\_FSE23}}.

\subsection{Experimental Configuration}
GCN and GAT use two message-passing layers; the one-layer ablation is examined separately in Section~\ref{sec:rq3}. Both use 256 hidden channels. GAT concatenates four attention heads with 64 channels per head at each layer. The MLP has one hidden layer with 256 units. GCN and GAT apply dropout of 0.01 to node activations after each ReLU; the MLP uses no dropout. All three encoders are followed by the same kind of linear classification head. Thus, the encoders share the features, classifier form, and self-training schedule, but are not matched in depth or parameter count. All three models are trained with the Adam optimizer at a learning rate of 0.001, minimizing cross-entropy loss over the architectural modules on the labeled and pseudo-labeled files.

All models use the same iterative learning procedure: a warm-up phase of $E_w = 100$ epochs on the initial labeled set, followed by at most $R = 4$ self-training rounds of $E_s = 30$ epochs each. The initial labeled set contains 5\% of a system's files, bounded between 30 and 200 files. The lower bound provides sufficient supervision in small systems, where 5\% alone would yield too few labeled files, while the upper bound limits the manual labeling effort in large systems such as Chrome. After sampling, we add one randomly selected file from each module still missing from the initial labeled set, so every module has at least one labeled file before training. During self-training, only predictions with confidence above $\tau = 0.95$ are added as pseudo-labels. These settings are fixed across the neural models. We use the first 100 supplied fixed splits, with run IDs 0--99, for all methods and ablations. The initial 200-file cap applies before adding representatives of missing modules; the final seed set can therefore exceed 200 files.

The neural models were trained on an NVIDIA GB10 GPU using Python 3.11, PyTorch 2.9.1, and PyTorch Geometric 2.7.0; NBA ran on the CPU. Node features were generated once per system and reused across runs and neural configurations. The training seed was set to $42+r$ for split $r$, independently of feature generation. The complete software environment, configurations, and split identities are recorded in the replication package.

To examine sensitivity to initialization and training randomness, we also conducted an exploratory follow-up using the first ten splits crossed with five training seeds, keeping the features fixed. The selected comparisons cover close model rankings and depth or direction effects, with the large Chrome graph advantage included as a reference. A subset was repeated under deterministic GPU execution after a same-seed discrepancy was observed on A.UML. These follow-ups are reported separately from the primary 100-run experiment.
\subsection{Baseline: Naive Bayes\label{sec:baselines}}
We compare against \emph{NBAttract}~\cite{olsson2022automatically} (NBA), a multinomial Naive Bayes classifier originally proposed for semi-supervised code-to-architecture mapping. NBA represents each entity using text features extracted from code identifiers and package names, combined with module-level Concrete Dependency Abstraction (CDA) sentences, document terms that encode cross-module structural relations, e.g., \emph{A-Inherits-B} when an entity in module $A$ inherits from one in module $B$. Olsson et al.\ report that NBA outperforms \emph{CountAttract}~\cite{christl2007automated} and the information retrieval methods \emph{IRAttract} and \emph{LSIAttract}~\cite{bittencourt_IR} on the same benchmark~\cite{olsson2022automatically}, so we treat NBA rather than those earlier methods as the reference point for prior work. 

NBA was originally proposed and evaluated on Java systems only; we extend it to C/C++ systems in our benchmark and re-implement the algorithm in Python, following the original iterative procedure, including its 0.90 confidence threshold for adding pseudo-labels and the same initial labeled set used by the other methods. Self-training stops either when no remaining entity can be mapped with sufficient confidence or when all entities have been mapped. Code for our NBA re-implementation and C/C++ extension is available in the GitHub repository~\ref{fn:code}.
For the Python baseline, Java node names retain the qualified entity names supplied by the extractor; for C/C++, directory-qualified file base names provide the analogous naming context. Identifier tokens use English Snowball stemming. We use binary term occurrences, multinomial Naive Bayes with Laplace smoothing ($\alpha=1$), and class priors from the currently mapped files. Architectural module names also contribute training documents. This is a Python adaptation using scikit-learn, not a claim of numerical equivalence to the original Java/Weka implementation.
\subsection{Performance Evaluation}
Module distributions in the benchmark systems are highly imbalanced, with several modules containing only one or two files. We use macro-F1 to expose performance on smaller modules, which overall accuracy can obscure. We also report micro-F1, which equals accuracy for this single-label, exhaustive classification task. F1 is the harmonic mean of precision and recall. Precision measures the proportion of files assigned to a module that actually belong to it, whereas recall measures the proportion of files that truly belong to a module and are correctly assigned.

We report macro- and micro-averaged F1 scores. Macro-F1 is the unweighted average of per-module F1 scores over modules represented in the evaluated ground-truth subset. A module whose files all belong to the seed set has no held-out examples and is excluded from that macro average. Undefined precision and recall are assigned zero. Micro-F1 considers all predictions, including predictions into modules absent from the test ground truth, and reflects overall file-level mapping performance.

\section{Results\label{sec:results}}
We report results from 100 runs per model and system, using the same fixed labeled sets for all methods and ablations. All primary metrics cover the full test set, excluding the initial labeled files. For the neural models, each test file receives the final model's prediction, whether or not it was confidently mapped during self-training. NBA retains accepted assignments and assigns the remaining files using their last attraction scores. Thus, no method benefits from evaluation on a smaller subset. Results are summarized as the median across the 100 runs. This section addresses the three research questions in order.

We use two-sided Wilcoxon signed-rank tests~\cite{wilcoxon} to assess statistical significance, pairing runs by their shared split ID. The compared neural configurations also share the training seed within each split. We apply Holm--Bonferroni correction at $\alpha = 0.05$, grouping comparisons by the question being tested: neural methods (60 tests), neural methods against NBA (60), relation types (40), depth (40), and direction (80). Each family includes both metrics and all ten systems. The tables report differences between the two medians, not medians of paired differences. Cliff's $\delta$~\cite{cliff} describes the distributional effect size, classified by magnitude as negligible ($|\delta| < 0.147$), small ($0.147 \leq |\delta| < 0.33$), medium ($0.33 \leq |\delta| < 0.474$), or large ($|\delta| \geq 0.474$)~\cite{romano2006}.
\phantomsection
\subsection*{RQ1. Overall Effectiveness of C2A Mapping \label{sec:rq1}}
To answer RQ1, we compare \nba{}, MLP, GCN, and GAT over 100 runs per system, all initialized from the same labeled set; \nba{} follows its own self-training procedure (Section~\ref{sec:baselines}), while MLP, GCN, and GAT share the iterative learning loop of Section~\ref{sec:iterative}. Table~\ref{tab:main_results} reports median macro and micro F1. All three neural methods outperform the Python NBA adaptation on both metrics across all ten systems, with statistically significant differences, although the size of the advantage varies. For example, MLP exceeds NBA by only 2.1 micro-F1 points on JabRef. We focus the remaining comparison on MLP, GCN, and GAT, with MLP assessing the contribution of node features without exploiting the dependency graph.

\begin{table*}[pos=t]
\centering
\caption{Median F1 scores (\(\times 100\)) across 100 runs. MA and MI denote macro- and micro-F1, respectively. The highest unrounded median for each system--metric pair is highlighted; highlighting does not imply statistical significance.\label{tab:main_results}}
\renewcommand{\arraystretch}{1.1}
\setlength{\tabcolsep}{4pt}
\begin{adjustbox}{max width=\textwidth}
\begin{tabular}{l |*{9}{cc|} cc}
\toprule
& \multicolumn{2}{c|}{A.UML}
& \multicolumn{2}{c|}{C.Img}
& \multicolumn{2}{c|}{JabRef}
& \multicolumn{2}{c|}{Lucene}
& \multicolumn{2}{c|}{SH-3D}
& \multicolumn{2}{c|}{T.Mates}
& \multicolumn{2}{c|}{Bash}
& \multicolumn{2}{c|}{HDF}
& \multicolumn{2}{c|}{Chrome}
& \multicolumn{2}{c}{HDC} \\
\cmidrule(lr){2-3}\cmidrule(lr){4-5}\cmidrule(lr){6-7}\cmidrule(lr){8-9}\cmidrule(lr){10-11}\cmidrule(lr){12-13}
\cmidrule(lr){14-15}\cmidrule(lr){16-17}\cmidrule(lr){18-19}\cmidrule(lr){20-21}
Method
& MA & MI
& MA & MI
& MA & MI
& MA & MI
& MA & MI
& MA & MI
& MA & MI
& MA & MI
& MA & MI
& MA & MI \\
\midrule
\nb{\nba{}} & \nb{32.6} & \nb{64.7} & \nb{17.5} & \nb{47.7} & \nb{47.2} & \nb{89.7} & \nb{60.6} & \nb{82.3} & \nb{39.6} & \nb{73.5} & \nb{32.5} & \nb{68.8} & \nb{27.6} & \nb{54.0} & \nb{33.4} & \nb{48.8} & \nb{9.5} & \nb{54.4} & \nb{44.0} & \nb{76.7} \\
MLP & \best{82.8} & 89.6 & 90.3 & 93.6 & 66.1 & 91.7 & 85.3 & 95.0 & \best{84.8} & 94.9 & 92.4 & 95.4 & 64.7 & 82.3 & \best{89.5} & \best{90.1} & 59.2 & 72.4 & \best{76.4} & 86.7 \\
GCN & 78.8 & 91.3 & 90.7 & 93.8 & 80.0 & 94.1 & 85.5 & 93.7 & 78.5 & 91.8 & 79.5 & 91.3 & 68.2 & 87.5 & 79.6 & 82.9 & 78.3 & 88.5 & 70.8 & \best{88.8} \\
GAT & 82.5 & \best{93.2} & \best{91.5} & \best{95.5} & \best{85.1} & \best{95.8} & \best{89.8} & \best{95.1} & 84.7 & \best{95.6} & \best{94.7} & \best{97.1} & \best{69.4} & \best{87.9} & 81.9 & 83.6 & \best{81.1} & \best{89.5} & 74.5 & 88.6 \\
\bottomrule
\end{tabular}
\end{adjustbox}
\end{table*}
Using unrounded medians, GAT achieves the highest macro F1 on six systems and micro F1 on eight. MLP has the highest macro F1 on A.UML, SH-3D, HDF, and HDC, although several differences are small. On micro F1, MLP leads on HDF and GCN on HDC. HDF and HDC are among the smallest and least connected benchmark systems, whereas SH-3D is comparatively well connected but has low modular cohesion; in these systems, MLP either outperforms GAT or remains competitive. On Chrome, the only large-scale system in the benchmark, GAT and GCN exceed MLP by 21.8 and 19.0 macro-F1 points, and by 17.1 and 16.1 micro-F1 points, respectively. Because Chrome is the only large-scale system, this result alone cannot establish whether the graph advantage generalizes as system size increases. GCN consistently outperforms NBA but does not occupy a consistent position between MLP and GAT. RQ2 examines these system-level differences.

The same encoder and training settings are used for Java and C/C++ systems, with the language-dependent naming and path handling described in the feature construction. Bash, HDF, and HDC are the smallest and sparsest systems in the C/C++ subset, and MLP is competitive with or outperforms the GNNs on two of them.

\srq{Statistical significance of observed differences.}
Table~\ref{tab:result_stats} reports pairwise statistical comparisons of the three learned methods for the performance differences shown in Table~\ref{tab:main_results}.
\begin{table}[pos=htbp]
\centering
\caption{Differences of median F1 scores in points for macro-F1 (MA) and micro-F1 (MI), with Cliff's $\delta$ magnitude in parentheses (N/S/M/L). Positive values favor the first method. Cells in which the second method is significantly better are highlighted. $\dagger$ marks a difference that is not statistically significant.\label{tab:result_stats}}

\renewcommand{\arraystretch}{1.15}
\setlength{\tabcolsep}{3pt}
\begin{adjustbox}{max width=\columnwidth}
\begin{tabular}{l | rr | rr | rr}
\toprule
& \multicolumn{2}{c|}{GAT $-$ GCN}
& \multicolumn{2}{c|}{GCN $-$ MLP}
& \multicolumn{2}{c}{GAT $-$ MLP} \\
\cmidrule(lr){2-3}\cmidrule(lr){4-5}\cmidrule(lr){6-7}
System & MA & MI & MA & MI & MA & MI \\
\midrule
A.UML & $+3.7$ (M) & $+1.9$ (L) & \best{$-4.0$ (M)} & $+1.7$ (S) & $-0.3^\dagger$ (N) & $+3.7$ (L) \\
C.Img & $+0.7$ (S) & $+1.7$ (L) & $+0.4^\dagger$ (N) & $+0.2^\dagger$ (N) & $+1.1$ (S) & $+1.9$ (L) \\
JabRef & $+5.0$ (M) & $+1.7$ (L) & $+13.9$ (L) & $+2.3$ (L) & $+19.0$ (L) & $+4.0$ (L) \\
Lucene & $+4.3$ (M) & $+1.4$ (M) & $+0.3^\dagger$ (N) & \best{$-1.3$ (M)} & $+4.5$ (M) & $+0.0^\dagger$ (N) \\
SH-3D & $+6.2$ (M) & $+3.8$ (L) & \best{$-6.3$ (L)} & \best{$-3.1$ (L)} & $-0.2^\dagger$ (N) & $+0.7^\dagger$ (N) \\
T.Mates & $+15.2$ (L) & $+5.9$ (L) & \best{$-12.9$ (L)} & \best{$-4.1$ (L)} & $+2.3$ (S) & $+1.8$ (M) \\
Bash & $+1.2^\dagger$ (N) & $+0.4^\dagger$ (N) & $+3.5^\dagger$ (S) & $+5.2$ (L) & $+4.7$ (S) & $+5.6$ (L) \\
HDF & $+2.2$ (S) & $+0.7^\dagger$ (N) & \best{$-9.8$ (L)} & \best{$-7.2$ (L)} & \best{$-7.6$ (L)} & \best{$-6.5$ (L)} \\
Chrome & $+2.8$ (L) & $+1.1$ (S) & $+19.0$ (L) & $+16.1$ (L) & $+21.8$ (L) & $+17.1$ (L) \\
HDC & $+3.6$ (S) & $-0.2^\dagger$ (N) & \best{$-5.5$ (S)} & $+2.0^\dagger$ (N) & $-1.9^\dagger$ (N) & $+1.9^\dagger$ (N) \\
\bottomrule
\end{tabular}
\end{adjustbox}
\end{table}
GAT significantly outperforms GCN on macro F1 for nine systems and on micro F1 for seven, with effects ranging from small to large. Neither difference is significant on Bash; on HDC, only the macro-F1 advantage is significant. GCN's difference from MLP varies substantially by system and metric. MLP significantly outperforms GCN on macro F1 on A.UML, SH-3D, T.Mates, HDF, and HDC, whereas GCN significantly outperforms MLP on JabRef and Chrome on both metrics. On Bash, GCN does not significantly outperform MLP on macro F1 but does on micro F1, with a large effect.

GAT significantly outperforms MLP on macro F1 for C.Img, JabRef, Lucene, T.Mates, Bash, and Chrome, and on micro F1 for A.UML, C.Img, JabRef, T.Mates, Bash, and Chrome. The differences on SH-3D are not significant on either metric. In contrast, MLP significantly outperforms GAT on both metrics only for HDF, with large effects. Despite its low connectivity, Bash favors GAT on both metrics; this exception is discussed in \hyperref[sec:rq2]{RQ2}.

\begin{figure}[pos=h]
    \centering
    \includegraphics[width=0.98\columnwidth]{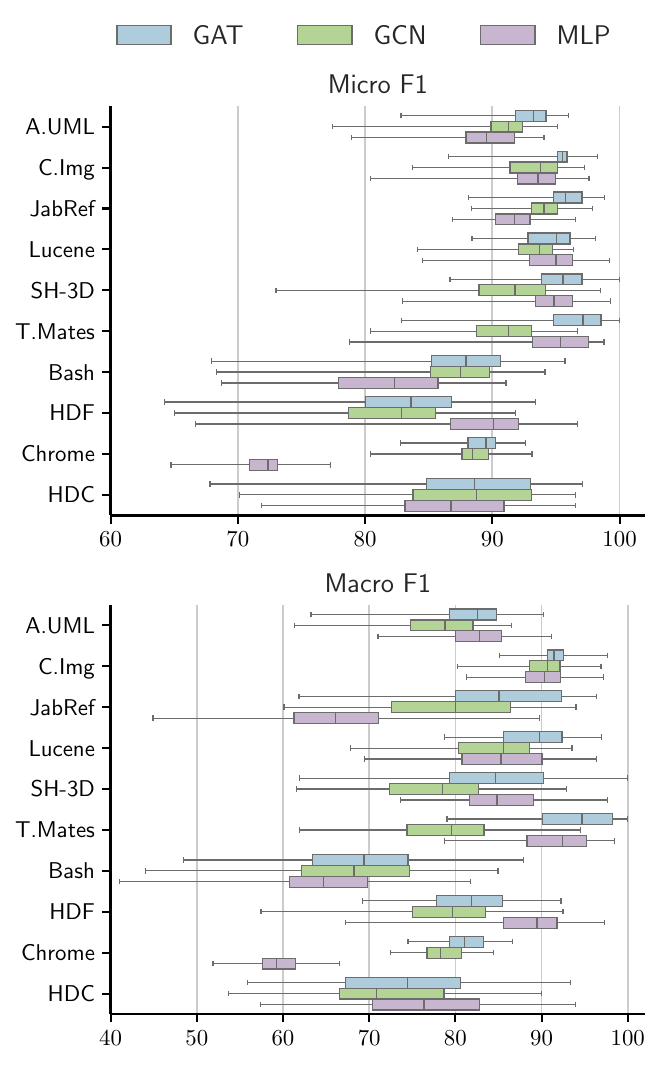}
    \caption{Distribution of micro F1 (top) and macro F1 (bottom), in points, across 100 runs for each model and system. Boxes show the median and interquartile range; whiskers span the observed range.\label{fig:rq1_variability}}
\end{figure}

\srq{How stable is mapping across runs?}
Each of the 100 runs uses a different labeled partition and training seed; Figure~\ref{fig:rq1_variability} therefore shows F1-score variability arising from both sources. Micro F1 is generally stable, with interquartile ranges below 6 points for seven of ten systems. Macro F1 varies more because changes in the initial labeled set disproportionately affect small modules. This pattern is shared across MLP, GCN, and GAT. HDC is the most variable system, with macro-F1 interquartile ranges of 13--14 points, compared with about 8 points for micro-F1.

Variability is associated with the number of initially labeled files and, for some systems, graph connectivity, rather than system size or language. HDC, HDF, Bash, and SH-3D start near the 30-file floor, leaving only one or two labeled files per module. HDC, HDF, and Bash are also sparse, further limiting the structural signal available to graph-based models, whereas SH-3D is comparatively well connected. C.Img provides a contrast: despite its small size, it is dense and cohesive and remains stable. Once the initial training set reaches roughly 50 files, interquartile ranges are generally narrow regardless of scale; C.Img and Chrome are among the most stable systems on macro F1. Remaining macro-F1 variability is associated with module-size skew: systems with both very large and small modules, such as JabRef, retain wider distributions despite adequate supervision. \hyperref[sec:seed_robustness]{RQ3} fixes the labeled partition across repeated runs to isolate the contribution of the training seed.

\begin{GrayBox}{Answer to RQ1.}
All three neural methods outperform \nba{} on both metrics across all ten systems. GAT achieves the highest averages of system-level median macro F1 (83.5) and micro F1 (92.2), leading on six and eight systems, respectively. GCN improves over \nba{} but does not consistently outperform the graph-free MLP baseline. Run-to-run variability is largest on macro F1 for the smallest systems. \hyperref[sec:rq2]{RQ2} examines the graph contribution beyond lexical features, while \hyperref[sec:seed_robustness]{RQ3} isolates variability from the training seed.
\end{GrayBox}

\phantomsection
\subsection*{RQ2. Contribution of the Dependency Graph\label{sec:rq2}}
RQ1 shows that no learned method dominates throughout the benchmark. RQ2 examines what the dependency graph and GNNs contribute beyond lexical features: which system properties are associated with the benefit of message passing, and how often each model correctly classifies individual files across runs, a question median F1 cannot answer since it aggregates performance at the system level. All three neural models use identical lexical features and training loops, although their encoder architectures and capacities differ.

\srq{Which system properties are associated with where the graph helps?} Connectivity describes how much information is available from neighboring files; homophily measures how often dependencies connect files in the same ground-truth module~\cite{gnn_homophily}. Bash, HDC, and HDF have the lowest average typed out-degrees in the benchmark ($4.2$, $5.4$, and $8.1$, respectively). MLP remains competitive with GAT on HDC and significantly outperforms it on HDF, but GAT significantly outperforms MLP on Bash. Sparse connectivity alone therefore does not determine whether message passing helps. These systems also differ in homophily, module composition, and the supervision available per module.

On the low-homophily systems SH-3D ($\mathcal{H}=0.36$) and T.Mates ($\mathcal{H}=0.34$), GCN performs significantly worse than MLP with large effect sizes. GAT is more competitive: neither difference from MLP is significant on SH-3D, while T.Mates favors GAT by 2.3 macro-F1 and 1.8 micro-F1 points. One possible explanation is that attention can assign less weight to uninformative cross-module neighbors than degree-normalized aggregation. We did not measure this mechanism directly, however, and the crossed follow-up shows seed-dependent rankings on both systems. On HDF ($\mathcal{H}=0.35$), both GNNs perform significantly worse than MLP. Together, these observations are compatible with connectivity and homophily influencing the value of message passing, but do not establish a rule for selecting an encoder.

\begin{figure}[pos=!h]
    \centering
    \includegraphics[width=1.0\columnwidth]{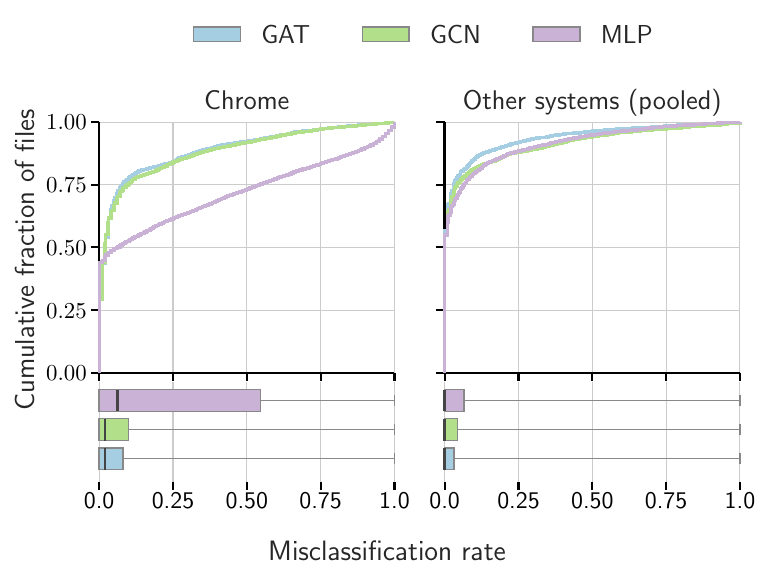}
    \caption{Empirical cumulative distribution of per-file misclassification rates across 100 runs, for GAT, GCN, and MLP, on Chrome (left) and pooled across the remaining nine systems (right). The y-axis is the cumulative proportion of test-set files whose misclassification rate is at or below the value on the x-axis, so a curve that rises earlier and higher indicates more reliable per-file decisions. The box plot below each panel shows the median, interquartile range, and full range of the misclassification rate for each model.\label{fig:rq2_combined}}
\end{figure}

\srq{Which files does the graph help most, and where does it fail?}
For each file, we compute its misclassification rate as the proportion of its test-set evaluations in which its predicted module differs from the ground-truth module, and refer to files with a rate of at least $0.9$ as \emph{persistently wrong}. Files that appear only in labeled sets are excluded. Figure~\ref{fig:rq2_combined} contrasts Chrome against the pooled distribution of the other nine systems, giving each evaluated file equal weight.

Across the nine small-to-medium-sized systems, the pooled distributions are concentrated near low error rates for all three models. Chrome shows a much larger contrast: MLP has a wider misclassification-rate distribution and a heavier tail of files misclassified in nearly every run. Among the 18,342 Chrome files evaluated at least once, 1,913 are persistently wrong under MLP (10.4\%), compared with 213 under GCN (1.2\%) and 204 under GAT (1.1\%). Thus, the graph advantage on Chrome is visible not only in aggregate F1 but also in the frequency of repeated file-level errors.

Across all ten systems, 241 files meet the persistent-error criterion under default GAT, of which 204 are in Chrome. These counts describe the current 100-run experiment; they can change with the labeled sets, training seeds, and error threshold. RQ3 examines how the persistent-error counts change when edge direction is altered.

\begin{GrayBox}{Answer to RQ2.}
Connectivity and homophily provide plausible context for the system-level differences, but neither alone determines whether a GNN improves on lexical features. GAT is more competitive than GCN on the low-homophily systems SH-3D and T.Mates, while MLP significantly outperforms both GNNs on HDF. On Chrome, both GNNs substantially improve aggregate F1 and reduce the share of persistently wrong files relative to MLP. These results support a contribution from dependency information, without isolating it from encoder capacity or establishing a causal model-selection rule.
\end{GrayBox}
\phantomsection
\subsection*{RQ3. Sensitivity to Design Choices\label{sec:rq3}}
RQ1 and RQ2 establish where the approach performs well and what the dependency graph contributes. RQ3 examines whether modeling dependency types separately is necessary and how sensitive the approach is to two message-passing design choices: the number of layers and the direction of aggregation. A crossed follow-up then tests whether the training seed alone can change the rankings observed in RQ1 and the ablation comparisons above. Because the benchmark systems have highly imbalanced module sizes, macro F1 can be volatile: it gives each module equal weight, so reassigning only a few files in a small module can substantially change the score. Micro F1 instead weights each file equally and may remain largely unchanged by the same reassignments. We therefore report both metrics throughout and regard a significant difference in either metric as evidence that the design choice affects performance.

\begin{figure}[pos=!tbp]
    \centering
    \includegraphics[width=1\columnwidth]{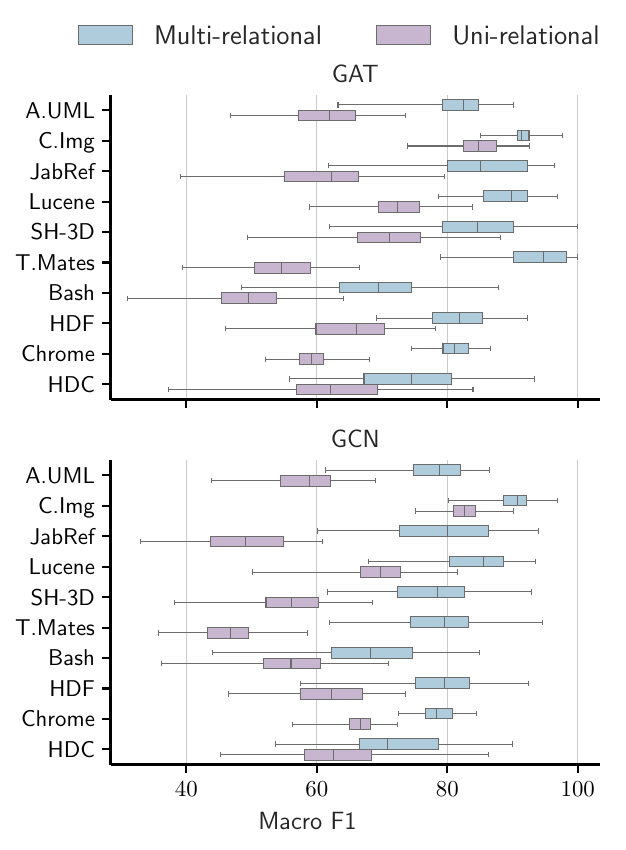}
    \caption{Distribution of macro F1, in points, across 100 runs for the multi-relational and uni-relational variants of GAT and GCN. Boxes show the median and interquartile range; whiskers span the observed range.\label{fig:homo_hetero}}
\end{figure}

\srq{Does modeling dependency types separately matter?} Our encoders keep one adjacency matrix per relation type and learn a separate transformation for each, rather than collapsing all dependencies into a single edge type. Figure~\ref{fig:homo_hetero} compares this multi-relational design against a uni-relational variant, in which all dependency types are merged into one undifferentiated adjacency matrix before message passing, for both GAT and GCN across the ten benchmark systems.
 
Both GAT and GCN perform significantly better with multiple relation types than with merged edges on every system and both metrics. Median macro-F1 gains range from 6.6 to 40.2 points for GAT and from 8.1 to 32.7 points for GCN. The smallest gain occurs on C.Img for both encoders, and the largest on T.Mates. This is consistent with different dependency types carrying different structural signals about module membership: a call edge and an inheritance edge need not carry the same evidence that two files share a module. The multi-relational encoder also has more parameters than the uni-relational one, so the experiment does not isolate relation typing from added capacity.

\srq{How many message-passing layers does the graph support?}
Additional GNN layers enlarge each node's receptive field, which is the $L$-hop neighborhood that can influence its representation after $L$ message-passing layers. However, when repeated aggregation mixes information from a large portion of the graph, node representations can become increasingly similar and lose the discriminative information needed for classification, an effect known as over-smoothing~\cite{gnn_oversmoothing}. We therefore explore only one and two layers, and Table~\ref{tab:depth} reports the comparison.
\begin{table}[pos=t]
\centering
\caption{\small Differences of median F1 scores in points for macro-F1 (MA) and micro-F1 (MI), with Cliff's $\delta$ magnitude in parentheses (N/S/M/L). Positive values favor two layers; cells in which one layer is significantly better are highlighted. $\dagger$ marks a difference that is not statistically significant. \label{tab:depth}}
\renewcommand{\arraystretch}{1.2}
\setlength{\tabcolsep}{8pt}
\begin{adjustbox}{max width=\columnwidth}
\begin{tabular}{l | rr | rr}
\toprule
& \multicolumn{2}{c|}{GAT (2L $-$ 1L)} & \multicolumn{2}{c}{GCN (2L $-$ 1L)} \\
\cmidrule(lr){2-3}\cmidrule(lr){4-5}
System & MA & MI & MA & MI \\
\midrule
A.UML & $+0.4^\dagger$ (N) & $-0.4^\dagger$ (N) & $-0.3^\dagger$ (N) & $-0.6^\dagger$ (N) \\
C.Img & $+0.1^\dagger$ (N) & $+0.0^\dagger$ (N) & $-0.4^\dagger$ (N) & \best{$-1.1$ (S)} \\
JabRef & $+3.2$ (S) & $+1.3$ (M) & $+5.1$ (M) & $+1.6$ (M) \\
Lucene & $+2.4^\dagger$ (N) & \best{$-1.1$ (S)} & $+2.8^\dagger$ (S) & $-0.5^\dagger$ (N) \\
SH-3D & $+0.9^\dagger$ (N) & $+0.7^\dagger$ (N) & $+0.0^\dagger$ (N) & \best{$-0.9$ (S)} \\
T.Mates & $+2.4$ (S) & $-0.3^\dagger$ (N) & \best{$-5.6$ (L)} & \best{$-2.6$ (L)} \\
Bash & $+5.0$ (M) & $+4.6$ (L) & $+6.1$ (S) & $+3.8$ (M) \\
HDF & $-2.0^\dagger$ (N) & $-0.9^\dagger$ (N) & $-0.1^\dagger$ (N) & $+1.6$ (N) \\
Chrome & $+2.9$ (L) & $+3.0$ (L) & $+0.1^\dagger$ (N) & $+1.3$ (L) \\
HDC & \best{$-6.0$ (S)} & $-0.5^\dagger$ (N) & \best{$-9.1$ (M)} & $-0.2^\dagger$ (N) \\
\bottomrule
\end{tabular}
\end{adjustbox}
\end{table}
According to Table~\ref{tab:depth}, two layers significantly improve GAT's macro F1 on JabRef, T.Mates, Bash, and Chrome, and its micro F1 on JabRef, Bash, and Chrome. They significantly reduce macro F1 on HDC ($-6.0$ points) and micro F1 on Lucene ($-1.1$ points). Neither difference is significant on SH-3D. For GCN, two layers significantly improve both metrics on JabRef and Bash, and micro F1 on HDF and Chrome. They significantly reduce micro F1 on C.Img and SH-3D, both metrics on T.Mates, and macro F1 on HDC. The HDF micro-F1 gain does not remain significant when all neural comparisons are corrected together.

These results support system-dependent depth choices. The gains on JabRef are compatible with informative two-hop neighborhoods, while the losses on T.Mates for GCN are compatible with greater mixing across modules. These mechanisms were not measured directly, however, and neither connectivity nor homophily alone explains every result. Two GAT layers benefit several systems in this benchmark, but are not uniformly preferable.

\srq{Does edge direction matter?} The heterogeneous encoder aggregates source-to-target by default, so a file's representation is shaped by the neighbors that depend on it rather than by the neighbors it depends on. Table~\ref{tab:ablation_direction} compares this against an undirected variant, which symmetrizes every edge, and a reverse variant, which aggregates target-to-source.
\begin{table*}[thbp]
\centering
\caption{\small Differences of median F1 scores in points for macro-F1 (MA) and micro-F1 (MI) under default directed (D) source-to-target aggregation, compared with UD (undirected, symmetrized edges) and RD (reversed, target-to-source), with Cliff's $\delta$ magnitude in parentheses (N/S/M/L). Positive values favor D; cells where the alternative is significantly better are highlighted. $\dagger$ marks a difference that is not statistically significant. \label{tab:ablation_direction}}
\renewcommand{\arraystretch}{1}
\setlength{\tabcolsep}{11pt}
\begin{adjustbox}{max width=\textwidth}
\begin{tabular}{l | rr | rr | rr | rr}
\toprule
& \multicolumn{4}{c|}{GAT} & \multicolumn{4}{c}{GCN} \\
\cmidrule(lr){2-5}\cmidrule(lr){6-9}
& \multicolumn{2}{c|}{D $-$ UD} & \multicolumn{2}{c|}{D $-$ RD} & \multicolumn{2}{c|}{D $-$ UD} & \multicolumn{2}{c}{D $-$ RD} \\
\cmidrule(lr){2-3}\cmidrule(lr){4-5}\cmidrule(lr){6-7}\cmidrule(lr){8-9}
System & MA & MI & MA & MI & MA & MI & MA & MI \\
\midrule
A.UML & $+2.7$ (S) & $+1.5$ (M) & \best{$-1.5$ (S)} & $-0.2^\dagger$ (N) & \best{$-0.9$ (S)} & \best{$-1.0$ (S)} & \best{$-3.3$ (M)} & \best{$-1.5$ (M)} \\
C.Img & $+2.9$ (L) & $+1.4$ (L) & $+5.8$ (L) & $+2.1$ (L) & $+3.8$ (M) & $+0.3^\dagger$ (S) & $+9.3$ (L) & $+1.1$ (S) \\
JabRef & $+10.7$ (L) & $+1.4$ (M) & $+8.3$ (L) & $+0.2^\dagger$ (N) & $+3.6^\dagger$ (N) & $-0.3^\dagger$ (N) & $+10.6$ (L) & \best{$-0.8$ (S)} \\
Lucene & $+3.1$ (M) & $+1.2$ (S) & \best{$-4.2$ (L)} & \best{$-1.2$ (M)} & $-2.0^\dagger$ (S) & \best{$-0.6$ (S)} & \best{$-7.4$ (L)} & \best{$-1.4$ (M)} \\
SH-3D & $+18.6$ (L) & $+7.4$ (L) & $+14.5$ (L) & $+5.9$ (L) & $+8.7$ (L) & $+3.6$ (L) & $+16.6$ (L) & $+5.2$ (L) \\
T.Mates & $+8.2$ (L) & $+1.4$ (S) & $+9.5$ (L) & $-0.1^\dagger$ (N) & $-0.9^\dagger$ (S) & \best{$-2.3$ (M)} & $+0.8^\dagger$ (N) & \best{$-4.8$ (L)} \\
Bash & $-1.6^\dagger$ (N) & $+0.4^\dagger$ (N) & $-1.5^\dagger$ (N) & $+0.1^\dagger$ (N) & \best{$-3.2$ (S)} & $-0.7^\dagger$ (N) & \best{$-3.7$ (S)} & $-0.4^\dagger$ (N) \\
HDF & $+3.6$ (M) & $+4.2$ (L) & $-3.0^\dagger$ (S) & $-1.5^\dagger$ (N) & $+3.1$ (S) & $+4.4$ (M) & \best{$-2.5$ (S)} & $+0.2^\dagger$ (N) \\
Chrome & $+6.2$ (L) & $+2.0$ (L) & $+4.3$ (L) & $-0.1^\dagger$ (N) & $+2.4$ (M) & $-0.1^\dagger$ (N) & $+2.6$ (L) & $-0.1^\dagger$ (N) \\
HDC & $+8.9$ (M) & $+1.3$ (N) & $+10.9$ (L) & $-0.2^\dagger$ (N) & $+3.1$ (S) & $+0.8^\dagger$ (N) & $+11.1$ (L) & $+1.8^\dagger$ (N) \\
\bottomrule
\end{tabular}
\end{adjustbox}
\end{table*}
As shown in Table~\ref{tab:ablation_direction}, source-to-target aggregation significantly outperforms the reversed version in macro F1 across six systems: C.Img, JabRef, SH-3D, T.Mates, Chrome, and HDC for GAT. Reversing the aggregation notably enhances macro F1 in A.UML and Lucene by 1.5 and 4.2 points, respectively, and raises micro F1 in Lucene by 1.2 points. No significant difference is observed between original and reversed aggregation on Bash or HDF. Compared to undirected aggregation, the directed default significantly improves at least one metric on nine systems, although no difference is significant on Bash.

GCN is more mixed: the default has a significant advantage over at least one alternative on six systems, and a significant disadvantage on six, with some systems in both groups. On JabRef, reversal reduces macro F1 but improves micro F1; the latter difference is not significant under the combined correction across all neural comparisons. These metric-dependent effects do not support one direction for every system.

At the file level, reversal reduces the number of persistently wrong GAT files from 11 to 6 on Lucene and leaves it at 9 on A.UML. Chrome illustrates a different trade-off: reversal reduces the count from 204 to 103 even though median macro F1 falls by 4.3 points; undirected aggregation raises the count to 322. Aggregate F1 and the size of the persistent-error tail can therefore move in different directions. These counts do not establish that the same individual errors are resolved without introducing new ones.

\srq{How robust are these rankings to the training seed?}\label{sec:seed_robustness}
The 100-run comparisons in RQ1 and the preceding ablations vary both the labeled partition and the training seed. We therefore conduct an exploratory follow-up to test whether changing only the training seed can alter model rankings. For each of ten fixed splits, we train each configuration with five seeds. We summarize each comparison using the mean paired difference, first averaging across seeds within each split. To quantify uncertainty, we use a crossed bootstrap that resamples splits and seeds separately. We apply a Bonferroni adjustment to the 95\% intervals for the 24 comparisons in the broader follow-up and retain this adjustment for the deterministic subset. Thus, the 50 runs per configuration are not treated as 50 independent splits.

Table~\ref{tab:seed_robustness} reports the selected comparisons repeated under deterministic execution. Even with the labeled files held fixed, the macro-F1 ranking changes across seeds on six of ten splits for GAT versus MLP on SH-3D, five for the same comparison on T.Mates, and eight for GAT versus GCN on HDC. None of these three intervals rules out the possibility of no difference. This does not overturn RQ1's significant advantages on T.Mates and HDC, but it shows those advantages are less consistent once the effect of seed is isolated. For the reversal comparison, Lucene holds up under this check, while A.UML does not. GAT's advantage over MLP on Chrome is the most robust result: it wins in all 50 split-seed combinations, with a large and consistent margin of about 22.6 points and an adjusted interval of [19.1, 26.6].

\begin{table}[pos=htbp]
\centering
\caption{Exploratory macro-F1 comparisons under deterministic execution. Effects are mean paired differences in F1 points, with family-adjusted 95\% crossed-bootstrap intervals. \emph{Flips} counts the number of splits in which both positive and negative differences occur across the five seeds. RD and D denote reversed and default directions.\label{tab:seed_robustness}}
\renewcommand{\arraystretch}{1}
\setlength{\tabcolsep}{6pt}
\begin{adjustbox}{max width=\columnwidth}
\begin{tabular}{llrrr}
\toprule
System & Comparison & Mean & Interval & Flips \\
\midrule
A.UML & GAT RD $-$ D & $+2.4$ & $[-1.9,\,6.4]$ & 3/10 \\
Lucene & GAT RD $-$ D & $+6.7$ & $[2.1,\,11.8]$ & 3/10 \\
SH-3D & GAT $-$ MLP & $+1.0$ & $[-5.2,\,7.0]$ & 6/10 \\
T.Mates & GAT $-$ MLP & $+2.1$ & $[-2.7,\,6.9]$ & 5/10 \\
Chrome & GAT $-$ MLP & $+22.6$ & $[19.1,\,26.6]$ & 0/10 \\
HDC & GAT $-$ GCN & $+4.5$ & $[-1.1,\,11.3]$ & 8/10 \\
\bottomrule
\end{tabular}
\end{adjustbox}
\end{table}

The deterministic follow-up was prompted by an A.UML run that changed when repeated with the same seed. Deterministic execution produced identical results in all twelve tested configuration-level repeat checks. However, it changed the selected mean comparisons by less than 0.1 F1 points and left all seed-ranking flip counts unchanged. Thus, the observed seed sensitivity cannot be explained by numerical nondeterminism alone.

\begin{GrayBox}{Answer to RQ3.}
Multi-relational aggregation improves both metrics on every system for both encoders, although part of the gain may reflect added capacity. Depth and direction are more system-dependent: two GAT layers benefit several systems, while reversed aggregation improves macro F1 on A.UML and Lucene. The deterministic follow-up supports the Lucene and Chrome advantages but shows that several closer rankings, on SH-3D, T.Mates, and HDC, depend on the training seed. No depth or direction is uniformly best, and direction changes can reduce persistent errors even when aggregate macro F1 decreases, as on Chrome.
\end{GrayBox}

\section{Threats to Validity\label{sec:threats}}

\parg{External Validity:} The ten benchmark systems span Java, C, and C++, but they do not represent the full range of architectural styles, application domains, development practices, or programming languages. They were selected because usable ground-truth mappings were available. Chrome is the only system at industrial scale; consequently, observations about scalability rest on a single system rather than a broader trend. None of the systems consists primarily of AI-generated code, so it remains unknown whether such code yields dependency structures and module cohesion comparable to those in the studied systems.

Our approach is transductive and assumes that the graph is fixed during training. Adding files or changing dependencies requires reconstructing the graph and retraining the model. The approach is therefore more suitable for periodic architecture analyses and maintenance workflows than for incremental mapping at commit time.

\parg{Internal Validity:} The dependency graph is derived through static analysis. Static extraction can miss dependencies that arise only at runtime, for example through reflection, function pointers, conditional compilation, or macro expansion. It can also introduce imprecise edges when references cannot be resolved unambiguously. Because message passing depends on the extracted graph, missing and incorrect edges can directly affect the model's predictions.

A further threat arises from version alignment between the analyzed source code and the published ground-truth mapping. Several benchmark mappings were created years before the source versions used in this study were obtained. In addition, the files included in a system can depend on its build configuration. The exact file set in the analyzed revision may therefore differ from that represented in the ground-truth mapping.

Files can plausibly belong to more than one module, and different annotators may assign such files differently. In older systems, architectural drift may also weaken the correspondence between a file and its assigned ground-truth module. Since the initial labeled set is sampled from the same mapping used for evaluation, an incorrect ground-truth assignment can both provide misleading supervision during training and penalize a prediction that is architecturally plausible.

The primary unit of analysis is the file, following the granularity of the available ground-truth mappings. A file that implements multiple concerns must nevertheless receive one module label. Neither the GNNs nor the compared baselines can represent multiple simultaneous module assignments.

The 100 initial labeled sets and training seeds are fixed and reused across neural configurations. Since both change between runs, the primary experiment does not separate their contributions to variability. The crossed follow-up holds each partition fixed while varying the training seed and reveals seed-dependent rankings on several systems. It covers only ten splits and five seeds, with comparisons selected after the main experiment, and should therefore be interpreted as exploratory rather than as a replacement for the primary analysis.

The primary experiment did not enforce deterministic GPU operations. Same-seed controls exposed a numerical discrepancy on A.UML that was amplified during training and pseudo-label selection. Deterministic execution produced identical results in the tested repeat controls, but did not remove the seed-dependent rankings or materially change the selected average comparisons. This check does not establish repeatability on every software or hardware configuration.

\parg{Construct Validity:} A fixed confidence threshold determines whether an unlabeled file receives a pseudo-label. This threshold captures how concentrated the model's softmax output is, rather than mapping quality alone. We evaluate predictions over the complete test set, preventing a model that pseudo-labels fewer files from benefiting from evaluation on an easier subset.

Directory paths are included in the node features, and in several benchmark systems the folder hierarchy closely reflects the reference module decomposition. Where this alignment is strong, performance may be optimistic because folder-path features provide partial information about the target module, rather than requiring the model to infer the mapping from identifiers and dependencies alone.

The encoders share features and a training schedule but differ in depth, dropout, and parameter count. In particular, the multi-relational GNNs have more parameters than their merged-relation counterparts. Their advantages cannot therefore be attributed solely to graph structure or relation typing. The NBA baseline is a Python adaptation extended to C/C++, not a numerically verified reproduction of the original Java implementation.

Statistical significance depends in part on the comparison family. Combining all neural comparisons into one correction family leaves most conclusions unchanged, but removes significance for GCN's micro-F1 depth gain on HDF and reversal gain on JabRef. Persistent-error counts also depend on the chosen threshold and the files' test-set appearances; they describe this experiment rather than a seed-invariant set of difficult files.

\section{Conclusion\label{sec:conclusion}}

We studied code-to-architecture mapping as semi-supervised node classification over a directed multi-relational dependency graph, comparing GAT and GCN against an MLP baseline that receives the same lexical features and passes through the same self-training loop. Across ten systems in Java and C/C++ and 100 runs per system, all three neural models outperform the Naive Bayes reference on both metrics. GAT achieves the highest averages of system-level median macro F1 (83.5) and micro F1 (92.2), although MLP remains competitive or superior on several systems. On Chrome, the largest system in the benchmark at 18,343 files, GAT and GCN outperform MLP by 22 and 19 macro-F1 points, respectively. With only one large-scale system in the benchmark, this gap cannot establish whether the graph's advantage scales with system size more generally.

Two system properties are plausibly associated with where this advantage appears: connectivity sets a baseline for how much neighborhood information a file can aggregate, and homophily measures how often dependency edges connect files in the same ground-truth module. These associations are observational and do not establish a causal selection rule. At the file level, Chrome again shows the clearest contrast: MLP persistently misclassifies 10.4\% of the files observed in a test set, compared with 1.2\% for GCN and 1.1\% for GAT.

Letting the encoder learn a separate transformation per relation outperforms a single merged edge type on every system, although this also increases model capacity. Depth and direction have more system-dependent effects. Two GAT layers benefit several systems but are not uniformly preferable, and reversed aggregation improves macro F1 on A.UML and Lucene. Direction can also trade aggregate performance against persistent errors: on Chrome, reversal reduces the persistent-error count despite a lower macro F1.

The crossed seed study supports the large graph advantage on Chrome and the reversal advantage on Lucene, while showing that several other rankings vary with the training seed. Deterministic execution removes discrepancies in the tested same-seed controls without resolving this sensitivity. Average advantages should therefore not be read as guarantees for an individual labeled set or training run.

Three directions could clarify and extend the contribution of dependency information. First, an additional attention mechanism over relation types could replace the unweighted sum of relation outputs when different dependencies provide conflicting evidence; the current GAT already learns separate neighbor-attention parameters for each relation. Second, the graph could incorporate finer-grained architectural information, including folder hierarchies and class members, to better reflect how architects conceptualize modules. Third, separating identifier-based lexical features from path-based features would clarify how much of the MLP baseline's signal comes from naming versus directory structure, and how much of that overlaps with what the dependency graph already encodes.

\printcredits

\section*{Declaration of use of AI-assisted technologies}
We used various AI tools to review the manuscript and the source code used for the experiments. We used the output from these tools to manually update the manuscript and fix any issues found in the source code. We take full responsibility for the content of the manuscript and the source code provided in the reproduction package.

\section*{Funding}
Rakhshanda Jabeen was supported by the Industrial Graduate School ``Data Intensive Applications,'' funded by the Knowledge Foundation, Sweden (grant no. 20190336).

The Knowledge Foundation had no role in the study design; collection, analysis, or interpretation of data; preparation of the manuscript; or decision to submit the article for publication.

\section*{Declaration of competing interest}
The authors declare that they have no known competing financial interests or personal relationships that could have appeared to influence the work reported in this paper.

\section*{Data Availability}
The replication package, including source code, experimental splits, and results, is available at \url{https://github.com/sse-lnu/GCN-SAEroCon25}, together with instructions for obtaining the benchmark data and reproducing the experiments. Upon acceptance, the version corresponding to this article will be archived on Zenodo, and its DOI will be included in the final manuscript.

\bibliographystyle{cas-model2-names}
\bibliography{References}

\end{document}